\documentclass[prb,groupedaddress,showpacs]{revtex4}
\usepackage[dvips]{graphicx}
\begin{document}
\title{Charge qubits in semiconductor quantum computer architectures: Tunnel
coupling and decoherence}
\author{Xuedong Hu}
\affiliation{Department of Physics, University at Buffalo, the State
University of New York, Buffalo, NY 14260-1500}
\author{Belita Koiller}
\affiliation{Instituto de F{\'\i}sica, Universidade Federal do Rio de
Janeiro, 21945, Rio de Janeiro, Brazil}
\author{S. Das Sarma}
\affiliation{Condensed Matter Theory Center, Department of Physics,
University of Maryland, College Park, MD 20742-4111}
\date{\today}

\begin{abstract}
We consider charge qubits based on shallow donor electron states in
silicon and coupled quantum dots in GaAs.  Specifically, we study the
feasibility of P$_2^+$ charge qubits in Si, focusing on single qubit
properties in terms of tunnel coupling between the two phosphorus donors and
qubit decoherence caused by electron-phonon interaction.  By taking into
consideration the multi-valley structure of the Si conduction band, we show
that inter-valley quantum interference has important consequences for
single-qubit operations of P$_2^+$ charge qubits.  In particular, the valley
interference leads to a tunnel-coupling strength distribution centered around
zero.  On the other hand, we find that the Si bandstructure does not
dramatically affect the electron-phonon coupling and consequently, qubit
coherence.  We also critically compare charge qubit properties for Si:P$_2^+$
and GaAs double quantum dot quantum computer architectures. 
\end{abstract}
\pacs{
71.55.Cn, 
03.67.Lx, 
85.35.-p  
}
\maketitle

\section{Introduction}

Among the solid state candidates for qubits in quantum information processing,
semiconductor-based systems have been among the most extensively explored. 
Key features in favor of these proposals are the high level of theoretical
understanding, experimental control, and nanofabrication capabilities
currently available for semiconductors.  It is commonly believed that
group-IV or III-V semiconductor nanostructure-based quantum computer
architectures should be relatively easily scalable because of the existence
of the vast semiconductor microelectronics infrastructure.  This scalability
incentive has led to a great deal of recent activities in studying qubit
properties of semiconductor nanostructures.\cite{Schladming,SSC_Cardona}
Theoretically, many semiconductor-based quantum computers were conceived
to rely on either electron spins or nuclear spins as
qubits.\cite{Exch,Kane,Vrijen,Imamoglu99,Kane1,larionov01,skinner03,overview} 
Spin-$1/2$ fermions (electrons or nuclei) probably constitute the most natural
and robust choices of quantum two-level systems for qubits in solids. 
Unfortunately, and in spite of considerable recent
progress,\cite{Rugar04,Elzerman} it is difficult to perform fast measurements
of single electron (or nuclear) spins, which are required for practical
quantum computing implementations incoporating quantum error corrections.  The
problem here is quantitative--although the electron spin is surely a quantum
two-level system, the Bohr magneton is a very small quantity, (and spin
usually does not couple strongly to external probes), making it difficult to
manipulate and measure the microscopic single spin states in the solid state
environment.  

In contrast to spin qubits, charge qubits in semiconductors have the
substantial advantage of being easy to manipulate and measure since the
experimental techniques for measuring single electron charges in
semiconductors are extremely well-developed.  The price one pays for the
relative ease in the manipulation and read-out of single-charge states is, of
course, the strong decoherence and the rather short decoherence time of the
orbital charge states because they couple strongly to the environment through
the long-range Coulomb interaction.  This fast decoherence of orbital states
makes semiconductor charge qubits rather unlikely candidates for a scalable
quantum computer architecture.  However, the strong interactions make the
orbital states an excellent choice for studying qubit dynamics and qubit
coupling in the solid state nanostructure environment.  This is particularly
true in view of the difficulties encountered in the manipulation and the
measurement of the single spin states in semiconductors.  It is worthwhile
also to remember that the much-studied superconducting-Cooper-pair-box-based
quantum computer architectures are charge-based systems as
well,\cite{Schon_RMP,Nakamura0} and there are conceptual and formal overlaps
between semiconductor charge qubits and superconductor charge qubits,
providing further impetus for studying orbital qubits in semiconductor
nanostructures.  

There have been several proposals for orbital/charge qubits in
semiconductors.\cite{Barenco,Ekert,Tanamoto,Sherwin,hollenberg1,hollenberg2} 
In this work we theoretically analyze single charge qubit properties for
phosphorus donor states in silicon, comparing it critically with charge qubit
states in coupled quantum dots (QD) in GaAs.  Our specific goal is to
investigate how the peculiar six-fold valley degeneracy of Si conduction band
affects the single qubit properties of orbital states in P-doped Si system. 
The issue is important in the context of our earlier results showing that
quantum interference between valleys leads to a strong suppression of the
exchange energy controlling the inter-qubit coupling in the
electron-spin-based silicon quantum computer architecture.\cite{KHD1,KHD2} 
The dramatic adverse effect of the valley interference on the silicon
exchange gate naturally raises the question of whether a similar valley
interference effect would also strongly (and adversely) affect the charge
qubit properties.  We answer this question in this paper.

Successful coherent manipulation of electron {\it orbital} states in GaAs has
been achieved for electrons bound to donor impurities \cite{Cole00} as well
as electrons in double quantum dots.\cite{Hayashi}  There were also
suggestions of directly using electron orbital states in GaAs or Si as the
building blocks for quantum information
processing.\cite{Ekert,Tanamoto,hollenberg1,hollenberg2}  Specifically, a
double QD with an electron bound in each dot or a pair of phosphorus donors
that sit relatively close to each other (so as to have sizable wave function
overlap) form an effective hydrogen molecule in GaAs or Si host material. 
Charge qubits may be defined by ionizing one of the bound electrons, thus
leading to a double well potential filled with a single electron: The single
electron ground state manifold, whether it is the two states localized in
each of the wells or their symmetric and anti-symmetric combinations, can
then be used as the two-level system forming a charge
qubit.\cite{Ekert,Tanamoto}  The advantage of such a charge qubit is that it
is easy to manipulate and detect, while its disadvantage, as already
mentioned above, is the generally fast charge decoherence as compared to
spin.  

Here we study the feasibility of the P$_2^+$ charge qubit in Si, focusing on
single qubit properties in terms of the tunnel coupling between the two
phosphorus donors (Sec.\ref{sec:gap}), and charge decoherence 
of this system in terms of electron-phonon coupling (Sec.\ref{sec:phonon}).  
We take into consideration the multi-valley 
structure of the Si conduction band and explore whether valley interference
could lead to potential problems or advantages with the operations of P$_2^+$
charge qubits, such as difficulties in the control of tunnel coupling similar
to the control of exchange in two-electron systems,\cite{KHD1,KHD2} or
favorable decoherence properties through vanishing electron-phonon coupling.
In section \ref{sec:discuss} we critically compare charge
qubits based on Si:P$_2^+$ and GaAs double QD systems.

\section{The symmetric-antisymmetric gap for the P$_2^+$ molecule in Silicon:
Qubit fidelity}
\label{sec:gap}

We study the simple situation where a single electron is shared by a donor
pair, constituting a P$_2^+$ molecule in Si.  The charge qubit here consists
of the two lowest energy orbital states of an ionized P$_2$ molecule in Si
with only one valence electron in the outermost shell shared by the two P
atoms.  The key issue to be examined is the tunnel coupling and the resulting
coherent superposition of one-electron states, rather than the entanglement
among electrons, as occurs for an exchange-coupled pair of electrons.  

The donors are at substitutional sites ${\bf R}_A$ and ${\bf R}_B$ in an
otherwise perfect Si structure.  In the absence of an external bias, and for
well separated donors, we may write the eigenstates for the two lowest-energy
states as a superposition of single-donor ground state wavefunctions localized
at each donor, $\psi_A ({\bf r})$ and $\psi_B ({\bf r})$, similar to the
standard approximation for the H$_2^+$ molecular ion.\cite{slater}  The
symmetry of the molecule leads to two eigenstates on this basis, namely the
symmetric and antisymmetric superpositions
\begin{equation}
\Psi_\pm({\bf r})=\frac{\psi_A({\bf r}) \pm \psi_B({\bf r})}{\sqrt{2(1 \pm
S)}}~,
\label{eq:wav}
\end{equation}   
where $S = \langle\psi_A|\psi_B\rangle$ is the overlap integral and is
real.\cite{KHD2} 
The conduction band of bulk Si has six degenerate minima $(\mu=1,\ldots,6)$,
located along the $\Gamma-$X axis of the Brillouin zone at $|{\bf k}_\mu|\sim
0.85(2\pi/{\rm a})$ from the $\Gamma$ point, where ${\rm a} = 5.43\,$\AA 
~is the Si lattice parameter.  Following Kohn-Luttinger effective mass
approximation,\cite{Kohn} the single-donor ground state wavefunctions are
written in terms of the six unperturbed Si band edge Bloch states $\phi_\mu =
u_\mu({\bf r}) e^{i {\bf k}_{\mu}\cdot {\bf r}}$.  For the donor at ${\bf
R}_A$, 
\begin{equation}
\psi_A ({\bf r}) = \sum_{\mu = 1}^6  \alpha_\mu F_{\mu}({\bf r}-
{\bf R}_A) \phi_\mu ({\bf r}, {\bf R}_A)
= \sum_{\mu = 1}^6  \alpha_\mu F_{\mu}({\bf r}-
{\bf R}_A) u_\mu({\bf r}) e^{i {\bf k}_{\mu}\cdot ({\bf r}-{\bf R}_A)}~,
\label{eq:sim}
\end{equation}
where the envelope functions centered at the donor site, $F_{\mu}({\bf r}-{\bf
R}_A)$, are deformed shallow donor effective mass 1S hydrogenic orbitals.  For
instance, for $\mu = z$, 
$F_{z}({\bf r}) = \exp\{-[(x^2+y^2)/a^2 + z^2/b^2]^{1/2}\}/\sqrt{\pi a^2 b}$.   
The expansion coefficients $\alpha_\mu$, which are also called valley
populations, are real.\cite{footvalley}  The effective Bohr radii $a$ and $b$
are variational parameters chosen to minimize $E_{A} = \langle\psi_{{\bf
R}_A}| H_{A} |\psi_{{\bf R}_A}\rangle$, leading to $a=25$ \AA~and $b=14$ \AA~
when recently measured effective mass values are used in the
minimization.\cite{KHD1}  The operator $H_A$ (note that in our notation the
single donor Hamiltonians $H_A$ and $H_B$ are equivalent) is the single-donor
Hamiltonian for the bound electron,\cite{KCHD} which includes the kinetic
energy, the Si periodic potential, the impurity screened Coulomb potential
centered at ${\bf R}_A$, and the valley-orbit effects, leading to $E_A \sim
-40$ meV, consistent with the experimentally observed value of 45 meV. 

The Hamiltonian for the singly ionized donor pair $P_2^+$ can then be written
as 
\begin{equation}
H = H_A -\frac{e^2}{\epsilon|{\bf r}-{\bf R}_B|}+\frac{e^2}{\epsilon|{\bf
R}_A-{\bf R}_B|}~, 
\label{eq:h}
\end{equation}
from which it is straightforward to obtain the expectation value of the single
qubit energy gap between the lowest-lying (symmetric and antisymmetric)
states $\Psi_\pm({\bf r})$ in (\ref{eq:wav}):
\begin{equation}
\Delta_{\rm S-AS} = <\Psi_-|H|\Psi_-> - <\Psi_+|H|\Psi_+> = 
\frac{2}{1-S^2}\sum_{\mu=1}^6{\delta}_{\mu}({\bf R}) \cos ({\bf k}_{\mu}\cdot
{\bf R})~,
\label{eq:delta}
\end{equation} 
where $\alpha_\mu = 1/\sqrt{6}$ for unstrained Si,\cite{footvalley}  
$\delta_\mu({\bf R}) = |\alpha_\mu|^2(s_\mu C_1 - v_\mu)$ and expressions for 
$S$, $C_1$, $s_\mu$ and  $v_\mu$, all of which are functions of $\bf{R} =
\bf{R}_A-\bf{R}_B$, are given in Ref.~\onlinecite{KHD2}.  For $|{\bf R}|\gg
a,b$, $S\ll 1$, and the amplitudes $\delta_\mu({\bf R})$ are monotonically
decaying functions of the interdonor distance ${\bf R}$.  Except for the
anisotropy, which is a consequence of the effective mass anisotropy in Si,
the dependence of $\delta_\mu$ on $|{\bf R}|$ is qualitatively similar to the
symmetric-antisymmetric gap in the  $H_2^+$ molecule, namely an exponential
decay with power-law prefactors.  The main difference here comes from the
cosine factors, which are related to the oscillatory behavior\cite{KCHD}  of
the donor wavefunction in Si arising from the Si conduction band valley
degeneracy and to the presence of two pinning centers. 

\begin{figure}
\includegraphics[width=4.1in]
{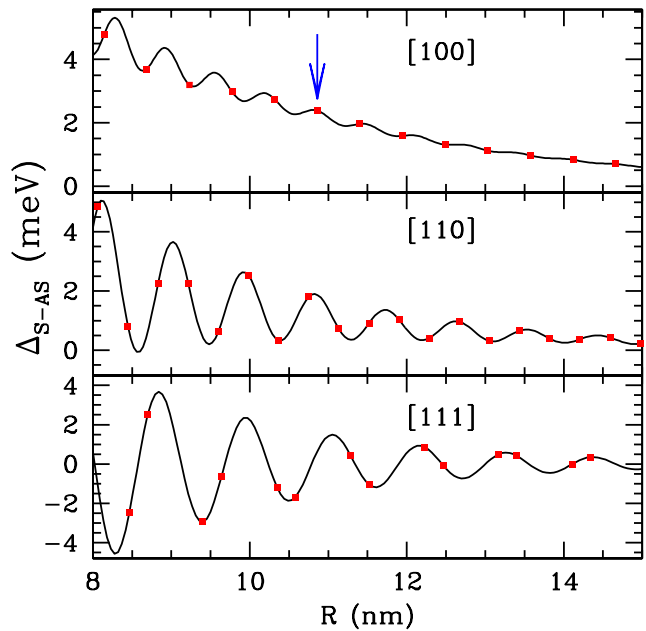}
\protect\caption[three frames with Delta vs R]
{\sloppy{
(color online) Symmetric-antisymmetric gap for the $P_2^+$ molecular ion in Si
for the donor pair along the indicated lattice directions. The arrow in the
upper frame indicates the {\it target} configuration analyzed in
Fig.~\ref{fig:distribution}.}}
\label{fig:delta}
\end{figure}

Figure~\ref{fig:delta} gives the calculated gaps as a function of ${\bf R}$
for a donor pair along three high-symmetry crystal directions.  
Two points are worth emphasizing here, which are manifestly different from the
corresponding hydrogenic molecular ion behavior: (i) $\Delta_{S-AS}$ is an
anisotropic and fast oscillatory function of ${\bf R}$; (ii) the sign of
$\Delta_{S-AS}$ may be positive or negative depending on the precise value of
${\bf R}$.  The characteristics mentioned in point (i) are similar to the
exchange coupling behavior previously discussed for the two-electrons neutral
donor pair.\cite{KHD1,KHD2,KCHD}  Point (ii) implies that the $P_2^+$
molecular ion ground state in Si may be symmetric (as in the $H_2^+$
molecular ion case) or antisymmetric depending on the separation between the
two P atoms.  Note that for the two-electron case, the ground state is always
a singlet (i.e. a symmetric two-particle spatial part of the wavefunction
with the spin part being antisymmetric), implying that the exchange $J$ is
always positive for a two-electron molecule.  For a one-electron ionized
molecule, however, the ground state spatial wavefunction can be either
symmetric or antisymmetric.     
\begin{figure}
\includegraphics[width=4.1in]
{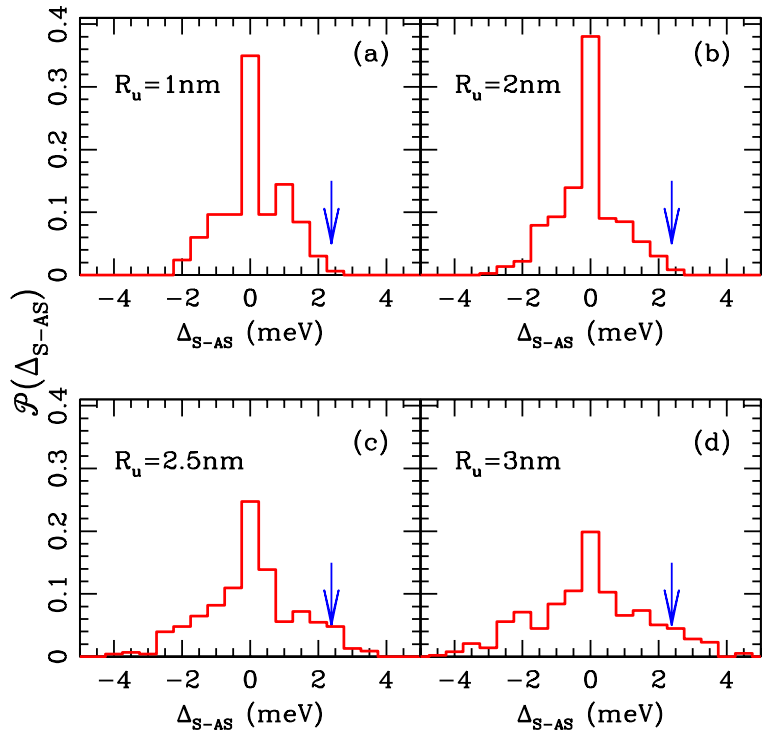}
\protect\caption[distributions of Delta]
{\sloppy{
(color online) Probability distributions of the symmetric-antisymmetric gap
for the $P_2^+$ molecular ion in Si.  Donor pairs are approximately aligned
along [100], but with an uncertainty radius $R_u$ with respect to this {\it
target} axial alignement (see text).  The arrow in each panel indicates the
gap value for the {\it target} configuration, for which the uncertainty
radius is $R_u=0$.  Notice that all four distributions are peaked at
$\Delta_{S-AS} =0$, and not at the {\it target} gap value. 
}}
\label{fig:distribution}
\end{figure}

Figure~\ref{fig:distribution} shows normalized probability distributions 
for the $\Delta_{S-AS}$ gap values when the first donor is kept fixed at ${\bf
R}_A$ and the second donor is placed at a site 20 lattice parameters away
($\sim$ 108.6 \AA), along the [100] axis.  This {\it target} configuration is
indicated by an arrow in Fig.~\ref{fig:delta}.  We allow the second donor
position ${\bf R}_B$ to visit all possible substitutional diamond lattice
positions within a sphere of radius $R_u$ centered at the attempted position. 
Our motivation here is to simulate the realistic fabrication of a P$_2^+$
molecular ion with fixed inter-atomic distance in Si with the state of the
art Si technology, in which there will always be a small ($R_u \sim 1-3$ nm)
uncertainty in the precise positioning of the substitutional donor atom
within the Si unit cell.  We would like to estimate the resultant randomness
or uncertainty in $\Delta_{S-AS}$ arising from this uncertainty in ${\bf
R}_B$.  For $R_u = 0$, i.e., for ${\bf R} = 20{\rm a} \hat x$, $\Delta_{S-AS}
\simeq 2.4$ meV, given by the arrows in Fig.~\ref{fig:distribution}.  We
incorporate the effect of small uncertainties by taking $R_u = 1$nm,
corresponding to the best reported degree of accuracy in single P atom
positioning in Si.\cite{encapsulation}  These small deviations completely
change the qubit gap distribution, as given by the histogram in
Fig.~\ref{fig:distribution}(a) strongly peaked around zero.  A similar
distribution is obtained for $R_u = 2$ nm, as illustrated in
Fig.~\ref{fig:distribution}(b).  Further increasing $R_u$ leads to broader
distributions of the gap values, though still peaked at zero [see
Fig.~\ref{fig:distribution}(c) and (d)].  This broadening is due to the fast
increase in the number of lattice sites inside the sphere of radius $R_u$,
thus contributing to the distribution, as $R_u$ increases.  We conclude that
the valley interference between the six Bloch states leads to a strong
suppression of the qubit fidelity since the most probable $\Delta_{S-AS}$
tends to be zero.

A very small $\Delta_{S-AS}$  is undesirable in defining the two states
$|0\rangle$ and $|1\rangle$ forming the charge qubit.  If we take them to be
the symmetric and anti-symmetric states given in Eq.~(\ref{eq:wav}), the fact
that they are essentially degenerate means that, when one attempts to
initialize the qubit state at $|0\rangle$, a different combination
$\alpha|0\rangle+\beta|1\rangle$ might result.  Well defined qubits may still
be defined under a suitable applied external bias, so that the electron
ground state wavefunction is localized around one of the donors, say at
lattice site ${\bf R}_A$, and the first excited state is localized arond
${\bf R}_B$.

Single qubit rotations, used to implement universal quantum
gates,\cite{NielsenChuang} might in principle be achieved by adiabatic
tunneling of the electron among the two sites under controlled axially
aligned electric fields through bias sweeps.\cite{Barrett}  When, at zero
bias, the ground state is {\it not} well separated by a gap from the first
excited state, severe limitations are expected in the adiabatic manipulation
of the electron by applied external fields.  In other words, the fidelity of
the single qubit system defining the quantum two-level dynamics will be
severely compromised by the valley interference effect.

\section{Electron-phonon coupling for a P$_2^+$ molecule}
\label{sec:phonon}

Two key decoherence channels for charge qubits in solids are background charge
fluctuations and electron-phonon coupling.\cite{Hayashi}  The former is
closely related to the sample quality (e.g., existences of stray charges and
charged defects in the system) and is extrinsic, while the latter is
intrinsic.  Here we focus on the electron-phonon coupling.  A critical
question for the P$_2^+$ molecular ion in Si is whether the Si bandstructure
and the associated charge density oscillations \cite{KCHD} lead to any
significant modification of the electron-phonon coupling matrix elements.  In
the following, we first derive the electron-phonon coupling for a single
valley situation, such as GaAs, to set a benchmark, then assess the effect of
the Si conduction band valleys and the Bloch functions on the donor
electron-phonon coupling matrix elements.  Our motivation is to investigate
whether valley interference leads to a strong suppression of the
electron-phonon coupling (similar to the suppression of exchange and tunnel
couplings), which would be beneficial for silicon charge qubits.

According to the results in Sec.\ref{sec:gap}, the energy splitting
$\Delta_{S-AS}$ between the two lowest energy states in a P$_2^+$ molecular
ion is up to a few meV, thus only low energy acoustic phonons near the
Brillouin zone center ${\bf q}\sim 0$ contribute to phonon-induced electron
decoherence.  Electron-acoustic-phonon coupling in a semiconductor can be
classified into two types: deformation potential and piezoelectric.  Since Si
is not polar, deformation potential is the only relevant interaction.  We
will thus focus on this interaction for the rest of this paper.  The
eletron-phonon interaction takes the form:\cite{Mahan}
\begin{equation}
H_{ep} = D \sum_{\bf q} \left( \frac{\hbar}{2 \rho_m V \omega_{\bf q}}
\right)^{1/2} |{\bf q}| \rho({\bf q}) (a_{\bf q}+a_{-\bf q}^{\dagger}) \,,
\end{equation}
where $D$ is the deformation constant, $\rho_m$ is the mass density of the
host material, $V$ is the volume of the sample, $a_{\bf q}$ and $a_{-\bf
q}^\dagger$ are phonon annihilation and creation operators, and $\rho({\bf
q})$ is the Fourier transform of the electron density operator: 
\begin{equation}
\rho({\bf q}) = \sum_{\lambda, \eta} c_{\lambda}^\dagger c_{\eta} \int d{\bf
r} \ e^{-i{\bf q} \cdot {\bf r}} \phi_{\lambda}^* ({\bf r}) \phi_{\eta} ({\bf
r}) \,,
\end{equation}
where $\lambda$ and $\eta$ are indices of electronic states/modes, $c_\lambda$
and $c_\lambda^\dagger$ are electronic annihilation and creation operators
for the $\lambda$-mode, while $\phi$ are mode functions.  For our two-donor
(or double-dot) situation, where we are only interested in the two lowest
energy single-electron eigenstates, we can choose them as the basis (so that
$\lambda$ and $\eta$ take the value of $+$ and $-$ as defined below) and the
electron-phonon coupling Hamiltonian is conveniently written in this
quasi-two-level basis in terms of the Pauli spin matrices $\sigma_x$ and
$\sigma_z$ (where spin up and down states refer to the two electronic
eigenstates):
\begin{eqnarray}
H_{ep} & = & D \sum_{\bf q} \left( \frac{\hbar}{2 \rho_m V \omega_{\bf q}}
\right)^{1/2} |{\bf q}| \left(A_r \sigma_x + A_{\varphi} \sigma_z \right)
\left( a_{\bf q}+a_{-\bf q}^{\dagger} \right) \,, \nonumber \\
A_r & = & \langle -|e^{i{\bf q}\cdot{\bf r}}|+\rangle \,, \nonumber \\
A_{\varphi} & = & \frac{1}{2} \left( \langle +|e^{i{\bf q}\cdot{\bf
r}}|+\rangle -
\langle -|e^{i{\bf q}\cdot{\bf r}}|-\rangle \right) \,.
\end{eqnarray}
Here the term proportional to $\sigma_x$ can lead to transition between the
two electronic eigenstates and is related to relaxation; while the term
proportional to $\sigma_z$ only causes energy renormalization of the two
electronic levels, but no state mixing, so that it only leads to pure
dephasing for the electronic charge states.

We first consider a double quantum dot with one electron (which is similar
to a singly ionized donor pair) in GaAs where the nondegenerate conduction
band minimum occurs at the $\Gamma$ point.  When the two dots or donors are
well separated and not strongly biased, the relevant single-electron states
are 
\begin{equation}
\Phi_+ = a\phi_A({\bf r}) + b\phi_B({\bf r})~;~  \Phi_-= b\phi_A({\bf r}) -
a\phi_B({\bf r}), 
\label{eq:gaas1}
\end{equation}
with $\phi_{A(B)}({\bf r}) = \varphi({\bf r} - {\bf R}_{A(B)})\, u_0({\bf
r})$, where $\varphi({\bf r})$ is a slowly varying envelope function, and the
Bloch function at the conduction band minimum (${\bf k}=0$ at $\Gamma$ point)
is equal to the periodic part $u_0({\bf r})$.  Though we have chosen the
envelopes $\varphi$ centered at each well to be identical, they could 
be different, as is generally the case for quantum dots.  

The deformation potential electron-phonon coupling matrix element for
relaxation between the initial unperturbed eigenstate $|-\rangle$ and the
final eigenstate $|+\rangle$ is proportional to $\langle -|e^{i{\bf q} \cdot
{\bf r}}|+\rangle$, where ${\bf q}$ is the phonon wavevector.  For a GaAs
double donor or double dot case the matrix element is proportional to
\begin{eqnarray}
A_r = \langle \Phi_-|e^{i{\bf q}\cdot{\bf r}}|\Phi_+\rangle & = & \int d{\bf
r} \ |u_{0}({\bf r})|^2 e^{i{\bf q} \cdot {\bf r}} \left\{ ab^* [\varphi
({\bf r})]^2 - a^*b [\varphi ({\bf r-R})]^2 \right. \nonumber \\
& & \left. +(|b|^2 - |a|^2) \varphi ({\bf r}) \varphi ({\bf r-R})\right\} \,.
\label{eq:gaas2}
\end{eqnarray}
For small energy splittings between the $\Phi_\pm$ states,
all terms in the integrand of Eq.~(\ref{eq:gaas2}) are slowly varying
functions in the interatomic spacing scale, except $|u_{0} ({\bf r})|^2$,
which is periodic and normalized in a primitive cell of the host material:
$\frac{1}{\Omega} \int_{\Omega} d{\bf r} |u_{0} ({\bf r})|^2 = 1$ where
$\Omega$ is the volume of the primitive cell.  This allows for the
approximation  
$\int d{\bf r}~ |u({\bf r})|^2 f({\bf r}) \approx \int d{\bf r}~ f({\bf
r})$,\cite{Bastard}
valid for slowly varying $f({\bf r})$, to be applied to (\ref{eq:gaas2}), 
leading to:
\begin{eqnarray}
A_r & = & (ab^* - a^*b e^{i {\bf q} \cdot {\bf R}}) \int d{\bf r} ~ e^{i{\bf
q} \cdot {\bf r}} [\varphi ({\bf r})]^2 \nonumber \\
& & + (|b|^2 - |a|^2) \int d{\bf r} ~ e^{i{\bf q} \cdot {\bf r}} \varphi ({\bf
r}) \varphi ({\bf r-R}) \,.
\label{eq:A_r}
\end{eqnarray}
Here the first integral is an on-site contribution modified by the phase
difference $e^{i {\bf q}\cdot {\bf R}}$ between the two dots/donors, while the
second integral is a two-dot contribution that is generally much smaller
because of the small overlap.

The dephasing matrix element $A_{\varphi}$ can be similarly calculated and the
result is
\begin{eqnarray}
A_{\varphi} & = & i \left(|b|^2 - |a|^2\right) e^{i{\bf q} \cdot {\bf R}/2}
\sin \frac{{\bf q} \cdot {\bf R}}{2} \int d{\bf r} ~ e^{i{\bf q}
\cdot {\bf r}} [\varphi ({\bf r})]^2 \nonumber \\
& & + (a^* b + ab^*) \int d{\bf r} ~ e^{i{\bf q} \cdot {\bf r}} \varphi ({\bf
r}) \varphi ({\bf r-R}) \,.
\label{eq:A_ph}
\end{eqnarray}
Notice that here the prefactors $|b|^2 - |a|^2$ and $a^* b + ab^*$ are for
on-site and off-site integrals, just the opposite to what we have in
Eq.~(\ref{eq:A_r}).  In other words, when $|b| \sim |a|$, $A_{\varphi}$ is
small, charge decoherence caused by electron-phonon interaction is dominated
by relaxation;\cite{Barrett,Fedichkin} when $|b|$ and $|a|$ are very
different (so that, for example, $|b| \sim 1$ and $|a| \sim 0$), charge
decoherence is dominated by pure dephasing.\cite{Fedichkin}  Below we will
focus on the relaxation matrix element $A_r=\langle -|e^{i{\bf q}\cdot{\bf
r}}|+\rangle$ as the contributing integrals are identical in the dephasing
matrix element $A_{\varphi}$.

We now consider a singly ionized phosphorus donor pair in Si, taking into
account the Si bandstructure.  For two donors not too close to each other,
and possibly detuned by an axially aligned electric field, the lowest energy
single-electron states are superpositions of $\psi_A$ centered at
${\bf R}_A$ [given in Eq.~(\ref{eq:sim})] and $\psi_B$ centered at ${\bf
R}_B$, similar to Eq.~(\ref{eq:gaas1}): 
\begin{equation}
\Psi_+ = a\psi_A({\bf r}) + b\psi_B({\bf r})~;~  
\Psi_- = b\psi_A({\bf r}) - a\psi_B({\bf r}), 
\label{eq:general}
\end{equation}
where the superposition coefficients $a$ and $b$ are of course not to be
confused with the effective Bohr radii.
If we choose ${\bf R}_A=0$, ${\bf R}_B = {\bf R}$, and $\Psi_\pm$ as the
initial and final states, the relaxation matrix element $A_r$ can be written
as
\begin{eqnarray}
\langle \Psi_-|e^{i{\bf q}\cdot{\bf r}}|\Psi_+\rangle & = & \sum_{\mu \nu}
\alpha_{\mu}^* \alpha_{\nu} \int d{\bf r} \ u_{\mu}^* ({\bf r}) u_{\nu} ({\bf
r}) e^{-i({\bf k}_{\mu} - {\bf k}_{\nu} - {\bf q}) \cdot {\bf r}} \left\{
ab^* F_{\mu} ({\bf r}) F_{\nu} ({\bf r}) \right. \nonumber \\
& & - a^*b F_{\mu} ({\bf r-R}) F_{\nu} ({\bf r-R}) e^{i({\bf k}_{\mu} - {\bf
k}_{\nu}) \cdot {\bf R}} + |b|^2 F_{\mu} ({\bf r}) F_{\nu} ({\bf r-R})
e^{-i{\bf k}_{\nu} \cdot {\bf R}} \nonumber \\
& & \left. - |a|^2 F_{\nu} ({\bf r}) F_{\mu} ({\bf r-R}) e^{i{\bf k}_{\mu}
\cdot {\bf R}} \right\} \,.
\label{eq:e-ph_ME}
\end{eqnarray}

If the energy splitting for the two double donor states is small (because
of large inter-donor separations or valley interference), the dominant
electron-phonon coupling is restricted to the regime of $|{\bf q}| \ll |{\bf
k}_{\mu}|$.  A simpler form of Eq.~(\ref{eq:e-ph_ME}) can then be obtained by
just keeping those integrals with $\mu = \nu$ (other integrals have fast
oscillatory integrands and are thus vanishingly small):
\begin{eqnarray}
\langle \Psi_-|e^{i{\bf q}\cdot{\bf r}}|\Psi_+\rangle 
& \cong & (ab^* - a^*b \ e^{i{\bf q}  \cdot {\bf
R}}) \sum_{\mu=1}^6 |\alpha_{\mu}|^2 \int d{\bf r} \ e^{i{\bf q} \cdot {\bf
r}} F_{\mu}^2 ({\bf r}) 
\nonumber \\
& & + (|b|^2 -|a|^2) \sum_{\mu=1}^6 |\alpha_{\mu}|^2 {\cos ({\bf k}_{\mu}
\cdot {\bf R}}) \int d{\bf r} \ e^{i{\bf q} \cdot {\bf r}} F_{\mu} ({\bf r})
F_{\mu} ({\bf r-R}) \,.
\label{eq:cosine}
\end{eqnarray}
Equation~(\ref{eq:cosine}) for double donor state in Si takes on quite similar
form as Eq.~(\ref{eq:A_r}) for double dot states in GaAs, with the first
sum in Eq.~(\ref{eq:cosine}) containing on-site contributions, and the second
sum containing off-site (inter-dot) contributions. 
Thus the first sum should generally outweigh the second even for $a \not\cong
b$.  However, if one of the coefficients $a$ or $b$ is very small, for
example due to electric field bias, the second sum may become dominant for
the relaxation matrix element.  However, as we mentioned above, in that
situation pure dephasing becomes the more important decoherence channel.  
In the case when the overlap integrals do make non-negligible contributions
(for example, when the two donors are detuned but not too strongly so, and
the two donors are sufficiently close so that the overlap integrals are not
vanishingly small), it is interesting to note that each of the terms in the
sum over the valleys is multiplied by the {\it same} ${\cos ({\bf k}_{\mu}
\cdot {\bf R}})$ factors which appear in Eq.~(\ref{eq:delta}).  The effect
again is to have results for the off-site contribution to the electron-phonon
coupling strongly oscillatory as a function of the interdonor
relative position ${\bf R}$.  The average overall effect, as illustrated in
Fig.~\ref{fig:distribution}, is to reduce the absolute value of the
relaxation coupling.

We now consider the possible contributions when $|{\bf q}|$ may not be 
negligibly small.  Indeed, in Si, for $\hbar \omega_{\bf q} \sim 5$ meV, $q
\sim 0.1
\frac{2\pi}{\rm a}$.  Thus, if the energy splitting between $\Psi_\pm$ states
is $\gtrsim 5$ meV, we need to include in our calculation phonon wave vectors
that may couple to
the periodic part of the Bloch functions as described
below.\cite{notedephasing}  Expanding the
periodic part of the Bloch functions $u_{\mu}$ in terms of
plane waves (restricted to the reciprocal lattice wave vectors) yields:
$$
u_\mu ({\bf r}) = \sum_{{\bf G}_\mu} C_{{\bf G}_\mu} e^{i{\bf G}_\mu \cdot
{\bf r}} \,,
$$
so that the relaxation matrix elements of Eq.~(\ref{eq:e-ph_ME}) becomes 
\begin{eqnarray}
\langle \Psi_-|e^{i{\bf q}\cdot{\bf r}}|\Psi_+\rangle & = &
\sum_{\mu,\nu} \sum_{{\bf G}_\mu,{\bf K}_\nu} \alpha_\mu \alpha_\nu C_{{\bf
G}_\mu}^* C_{{\bf K}_\nu} \int d{\bf r} \ e^{-i({\bf G}_\mu + {\bf k}_{\mu} -
{\bf K}_\nu - {\bf k}_\nu - {\bf q})\cdot {\bf r}} \nonumber \\
& & \times \left[ ab^* F_\mu ({\bf r}) F_\nu ({\bf r}) -a^*b F_\mu ({\bf r-R})
F_\nu ({\bf r-R}) e^{i ({\bf k}_\mu-{\bf k}_\nu) \cdot {\bf R}} \right.
\nonumber \\
& & \left. - |a|^2 F_\mu ({\bf r-R}) F_\nu ({\bf r}) e^{i {\bf k}_\mu \cdot
{\bf R}} + |b|^2 F_\mu ({\bf r}) F_\nu ({\bf r-R}) e^{-i {\bf k}_\nu \cdot
{\bf R}} \right] \,.
\label{eq:e-ph_mq}
\end{eqnarray}
Since $|{\bf q}|$ is always relatively close to zone center (i.e., $|{\bf
q}|$ is always much smaller than $2\pi/{\rm a}$), the largest contribution to
Eq.~(\ref{eq:e-ph_mq}) comes from terms with $\nu=\mu$ and ${\bf K}_\mu = {\bf
G}_\mu$.  These are the same terms that determine the matrix elements in the
small $\bf q$ limit, as given by Eq.~(\ref{eq:cosine}).  The important
question now is whether other terms will also contribute significantly 
when $|{\bf q}|$ is not particularly close to the zone
center.  Recall that more than 90\% of the spectral weight in $u_\mu$ comes
from five plane waves\cite{KCHD} (the rest of the $C_{\bf G}$ coefficients
are at least one order of magnitude smaller): For $u_x$, these are $G_x = 0,\
\frac{2\pi}{\rm a} (-1,\pm 1, \pm 1)$, so that $k_x + G_x \approx 0.85~
\frac{2\pi}{\rm a},~ \frac{2\pi}{\rm a} (-0.15, \pm 1, \pm 1)$ are the five
smallest wave vectors contributing to the Bloch function $u_x ({\bf r}) e^{i
k_x r_x}$.  There is thus one scenario when ${\bf G}_\mu + {\bf k}_{\mu} -
{\bf K}_\nu - {\bf k}_\nu - {\bf q}$ might have similar amplitude as ${\bf
q}$: when $\nu = - \mu$ and ${\bf G}_\mu - {\bf K}_\nu$ is parallel to ${\bf
k}_\mu$.  For example, for ${\bf q} = q\hat{x}$, there are terms with ${\bf
G}_x + {\bf k}_x - {\bf K}_{-x} - {\bf k}_{-x} = \frac{2\pi}{\rm a}(\pm 0.3,
0, 0)$.  Since these wave vectors correspond to wave lengths of the order of
15 \AA, while the donor effective Bohr radius is about 20 \AA, one needs to
carefully evaluate the integrals involving these terms as their oscillatory
integrands have the same length scale as the envelopes.  The
$\nu=-\mu$ contribution to the electron-phonon coupling matrix elements takes
the form (taking into consideration that $F_{-\mu} = F_{\mu}$, ${\bf
k}_{-\mu} = - {\bf k}_{\mu}$, and $\alpha_{-\mu} = \alpha_\mu$):
\begin{eqnarray}
\langle \Psi_-|e^{i{\bf q}\cdot{\bf r}}|\Psi_+\rangle_{\nu=-\mu} 
& = & \sum_{\mu = 1}^6 |\alpha_\mu|^2 \sum_{{\bf G}_\mu,{\bf K}_{-\mu}}
C_{{\bf
G}_\mu}^* C_{{\bf K}_{-\mu}} \nonumber \\
& & \times \left\{ [ab^* -a^*b \ e^{-i({\bf G}_\mu - {\bf K}_{-\mu} - {\bf q})
\cdot {\bf R}}] \right. \nonumber \\ 
& & \left. \times \int d{\bf r} \ e^{-i({\bf G}_\mu + {\bf k}_{\mu} -
{\bf K}_{-\mu} - {\bf k}_{-\mu} - {\bf q})\cdot {\bf r}} F_\mu^2 ({\bf r})
 + (|b|^2 - |a|^2) e^{i{\bf k}_\mu \cdot {\bf R}} 
\right. \nonumber \\
& & \left.
\times \int d{\bf r} \
e^{-i({\bf G}_\mu + {\bf k}_{\mu} - {\bf K}_{-\mu} - {\bf k}_{-\mu} - {\bf
q}) \cdot {\bf r}} F_\mu ({\bf r}) F_\mu ({\bf r - R}) \right\}.
\label{eq:e-ph_final}
\end{eqnarray}
For each $\mu$, only the 5 dominant ${\bf G}_\mu$ contributions mentioned
previously are included in the second summation above.  (As an example, in
Table~\ref{table_Bloch} we list data\cite{KCHD} for $\mu = \pm x$). 
For each ${\bf G}_\mu \neq 0$, there exists one and only one ${\bf K}_{-\mu}$
for which the exponent ${\bf G}_\mu + {\bf k}_{\mu} - {\bf K}_{-\mu} - {\bf
k}_{-\mu}$ is small.  If ${\bf G}_\mu = 0$, all exponents are large so that
the integrals should not be important in the sum in Eq.~(\ref{eq:e-ph_final}).
Without loss of generality, we consider the $\mu = x$ part of the sum, where
${\bf G}_x + {\bf k}_x - {\bf K}_{-x} - {\bf k}_{-x} = \frac{2\pi}{\rm a}
(-0.3, 0, 0)$.  Notice that this exponent is independent of which ${\bf G}_x$
is in consideration.  Therefore, all the terms except $C_{{\bf G}_\mu}^*
C_{{\bf K}_{-\mu}}$ can be factored out of the sum over ${\bf G}_\mu$ and
${\bf K}_{-\mu}$, so that the sum in Eq.~(\ref{eq:e-ph_final}) is
proportional to $\sum_{{\bf G}_\mu, {\bf K}_{-\mu}\not= 0 } C^*_{{\bf G}_\mu}
C_{{\bf K}_{-\mu}}$, which vanishes due to the symmetry of Si lattice, as
illustrated in Table~\ref{table_Bloch}.  Therefore, for intermediate $|{\bf
q}|$ the lowest order correction to the small $|{\bf q}|$ electron-phonon
coupling matrix element vanishes, so that Eq.~(\ref{eq:cosine}) is valid in
both small and intermediate $|{\bf q}|$ regimes for the phonons involved. 

\begin{table}[!hbt]
\begin{tabular}{|l| c c c c c|} \hline
${\bf G}_x$ & 0 & (-1,1,1) & (-1,1,-1) & (-1,-1,1) & (-1,-1,-1) \\
$C_{{\bf G}_x}$ & (0.343,0) & (-0.313, +0.313) & (-0.313, -0.313) & (-0.313,
-0.313) & (-0.313, 0.313) \\ \hline
${\bf K}_{-x}$ & 0 & (1,1,1) & (1,1,-1) & (1,-1,1) & (1,-1,-1) \\
$C_{{\bf K}_{-x}}$ & (0.343,0) & (-0.313, -0.313) & (-0.313, 0.313) & (-0.313,
0.313) & (-0.313, -0.313) \\ \hline
$C_{{\bf G}_x}^* C_{{\bf K}_{-x}}$ & (0.118,0) & (0,0.196) & (0,-0.196) &
(0,-0.196) & (0,0.196) \\ \hline
\end{tabular}
\protect\caption[Bloch coefficients]
{\sloppy{Here we give the 5 most important expansion coefficients for the
periodic part of the Bloch state $\phi_x$ and $\phi_{-x}$ [for example,
$\phi_x ({\bf r}) = \sum_{{\bf G}_x} C_{{\bf G}_x} e^{i({\bf G}_x + {\bf
k}_x) \cdot {\bf r}}$].\cite{KCHD}  Notice that $C_{{\bf G}_x}$ and $C_{{\bf
K}_{-x}}$ are complex in general. 
}}
\label{table_Bloch}
\end{table}

In summary, the electron-phonon coupling for a P$_2^+$ molecular ion
in Si formally behaves very similarly to that for a single electron
trapped in a GaAs double quantum dot (restricted to the deformation
interaction).  The more complicated multi-valley bandstructure of Si and the
strong inter-valley coupling introduced by the phosphorus donor atoms do not
cause significant changes in the electron-phonon coupling matrix elements. 
The only valley interference effect occurs when the overlap between the two
donors is not negligible.  Even then the interference among the valleys only
causes oscillatory suppression of the off-site contributions, which are
relatively small anyway.  Therefore, available
estimates\cite{Barrett,Fedichkin} of decoherence induced by electron-phonon
coupling based on a single-valley hydrogenic approximation in the P$_2^+$
system in Si should be valid.  In other words, the multi-valley quantum
interference effect does not provide any particular advantage (or
disadvantage) for single qubit decoherence in the Si:P donor charge-based
quantum computer architecture.

\section{Discussion and Summary}
\label{sec:discuss}

We have so far explored the feasibility of charge qubits based on the P$_2^+$ 
system in Si.  We find that this system possesses decoherence properties
(induced by electron-phonon coupling) similar to a GaAs double quantum dot.  
The Si bandstructure does, however, significantly (and adversely) influence
the tunnel coupling between the two phosphorus donors, so that for many
relative positions between the donors in a pair, the tunnel coupling becomes
quite small.  In other words, if two donors are randomly placed in a Si host,
keeping their distance approximately constant, the tunnel coupling between
the donor sites can vary over a wide range of values (peaked around zero)
because of the Si conduction band valley degeneracy.  This is obviously
rather bad news for charge qubits in the P$_2^+$ system in Si: It implies that
a large percentage of the fabricated charge qubits are unlikely to work
properly since energy splitting in these two-level systems is essential for
quantum computation.

The quasi-randomness of tunnel coupling in a P$_2^+$ donor molecular ion in Si
can be contrasted with the corresponding tunnel coupling in a double QD
in GaAs (or Si).  The Coulomb potential of a donor provides a natural strongly
localized confinement potential.  Thus donors are really {\it identical}
quantum dots, with fixed positions given by the donor nuclei, a fixed
effective Bohr radius, and a fixed ground state energy level relative to the
conduction band edge.  All donor qubits are therefore expected to have
identical properties except for the donor positioning problem.  The main
problem with donors in Si is that they break the local translational symmetry
and introduce a strong valley-orbit coupling.  Donor electron states are
therefore superpositions of Bloch states from all the conduction band edges. 
The valley interference effects are thus strong in a donor system such as
Si:P, leading to the atomic scale oscillations in two-electron exchange
studied before\cite{KHD1,KHD2} and single-electron tunneling studied here. 
On the other hand, a gated QD is a truly artificial atom, whose position,
shape, size, and energy levels are all determined by the applied gate
potentials on the metallic gates some 100 nm away.  The confinement produced
by gate potentials are generally quite smooth and shallow, and the barriers
between potential minima quite broad.  These slowly varying features of gated
QDs dictate that quantum dot electronic properties are very sensitive to the
tuning of applied gate potentials.  It is also inevitable that two QDs are
never identical even after careful calibration.  

\begin{figure}
\includegraphics[width=4.1in]
{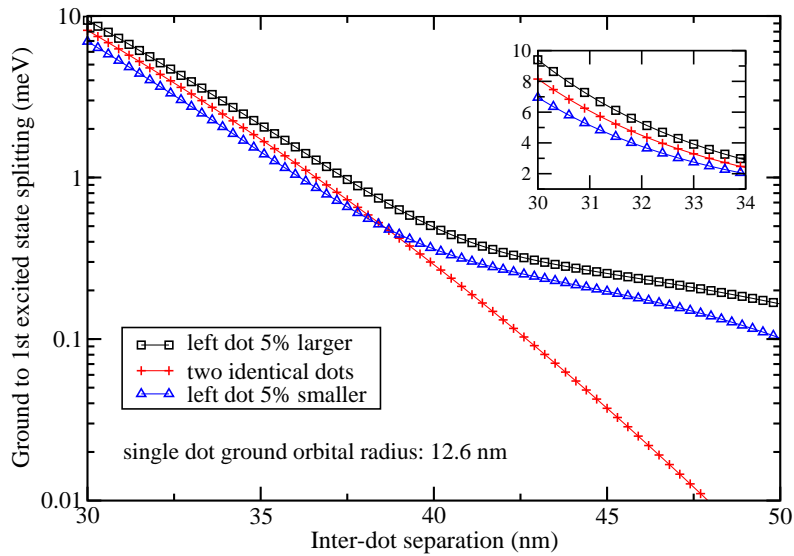}
\protect\caption[gap for uneven dots]
{\sloppy{
(color online) Energy gap between the ground and first excited states of a
single electron double dot as a function of inter-dot distance.  The crosses
are for two identical dots with a fixed Gaussian confinement\cite{HD} (with a
ground orbital radius of 12.6 nm); the squares are for situations where one
dot is 5\% larger; the triangles are for situations where that same dot is
5\% smaller.  At larger inter-dot distances the different-dot configurations
have larger energy splittings because the dot size difference introduces an
energy level detuning that is larger than the tunnel coupling.  The inset
plots the same data in the range of 30 to 34 nm inter-dot distance in the
linear scale, where we find that the 5\% dot size variation leads to a 10 to
20\% difference in the tunnel coupling.}}
\label{fig:uneven}
\end{figure}

As a simple illustration of the effect of uneven dots, we present in
Fig.\ref{fig:uneven} the dependence of tunnel coupling between a double
QD as a function of inter-dot distance.  We find that a 5\% dot
size variation leads to a 10 to 20\% difference in the tunnel coupling.  Note
that, in addition to size variation, there will be inevitable fluctuations in
inter-dot separations and barrier heights as well, leading to qubit
fluctuations.  Obviously, careful calibration is imperative for a QD system
to work as a reliable charge qubit.  In contrast, for donor-based charge
qubits, one does not need to be concerned with such dot size variation
problems since all P donors are identical.

In short, both donors in Si and gated QDs in either GaAs or Si pose difficult
challenges to solid state quantum information processing.  For donors the
challenge lies more on the fabrication process, while for gated dots the
challenge lies more on the gating control.  Which type of electron
confinement (carefully calibrated gated dots or carefully positioned donors)
may turn out to be better suited for quantum computing will ultimately be
determined by experimental work.

For electron decoherence we have so far limited ourselves to electron-phonon
coupling.  As we mentioned before, for charge degrees of freedom another
important source of decoherence is the fluctuation in charge traps close to
a charge qubit.\cite{Hayashi,Nakamura,Schon_RMP,Itakura}   Since charge
fluctuation noise can be treated in a very similar fashion as the
electron-phonon coupling,\cite{Schon_RMP} we do not anticipate any
significant qualitative difference between the Si:P system and GaAs quantum
dots.  Furthermore, since electron-phonon coupling is intrinsic, the
consequent decoherence is the limit that cannot be improved by having better
materials and fabrication quality.

In conclusion, we show that the inter-valley quantum interference leads to a
strong suppression of qubit fidelity in P$_2^+$ charge qubits in Si, as small
nanometer-scale fluctuations in donor positioning within the Si unit cell
produce an essentially random distribution (peaked around zero) in the energy
separation between the two levels defining the charge qubit.  We find
decoherence properties of charge qubits to be qualitatively unaffected by
multi-valley effects.  For QD-based charge qubits, we find variations in qubit
properties arising from fluctuations in dot sizes and separations, which will
have to be carefully characterized individually.

\begin{acknowledgments}

This work is supported by NSA, ARDA, ARO, and LPS at the University of
Maryland, by NSA, ARDA, and ARO at the University at Buffalo, and by Brazilian
agencies CNPq, FUJB, FAPERJ, PRONEX-MCT, and Instituto do Mil\^enio de
Nanoci\^encias-CNPq. 

\end{acknowledgments}

\bibliography{charge}

\begin{thebibliography}{39}
\expandafter\ifx\csname natexlab\endcsname\relax\def\natexlab#1{#1}\fi
\expandafter\ifx\csname bibnamefont\endcsname\relax
  \def\bibnamefont#1{#1}\fi
\expandafter\ifx\csname bibfnamefont\endcsname\relax
  \def\bibfnamefont#1{#1}\fi
\expandafter\ifx\csname citenamefont\endcsname\relax
  \def\citenamefont#1{#1}\fi
\expandafter\ifx\csname url\endcsname\relax
  \def\url#1{\texttt{#1}}\fi
\expandafter\ifx\csname urlprefix\endcsname\relax\def\urlprefix{URL }\fi
\providecommand{\bibinfo}[2]{#2}
\providecommand{\eprint}[2][]{\url{#2}}

\bibitem[{\citenamefont{Hu}(2004)}]{Schladming}
\bibinfo{author}{\bibfnamefont{X.}~\bibnamefont{Hu}}, \bibinfo{journal}{airXiv:
  cond-mat/0411012}  (\bibinfo{year}{2004}).

\bibitem[{\citenamefont{{Das Sarma} et~al.}(2004)\citenamefont{{Das Sarma}, {de
  Sousa}, Hu, and Koiller}}]{SSC_Cardona}
\bibinfo{author}{\bibfnamefont{S.}~\bibnamefont{{Das Sarma}}},
  \bibinfo{author}{\bibfnamefont{R.}~\bibnamefont{{de Sousa}}},
  \bibinfo{author}{\bibfnamefont{X.}~\bibnamefont{Hu}}, \bibnamefont{and}
  \bibinfo{author}{\bibfnamefont{B.}~\bibnamefont{Koiller}},
  \bibinfo{journal}{airXiv: cond-mat/0411755}  (\bibinfo{year}{2004}),
  \bibinfo{note}{to be published in Solid State Communications}.

\bibitem[{\citenamefont{Loss and DiVincenzo}(1998)}]{Exch}
\bibinfo{author}{\bibfnamefont{D.}~\bibnamefont{Loss}} \bibnamefont{and}
  \bibinfo{author}{\bibfnamefont{D.~P.} \bibnamefont{DiVincenzo}},
  \bibinfo{journal}{Phys. Rev. A} \textbf{\bibinfo{volume}{57}},
  \bibinfo{pages}{120} (\bibinfo{year}{1998}).

\bibitem[{\citenamefont{Kane}(1998)}]{Kane}
\bibinfo{author}{\bibfnamefont{B.~E.} \bibnamefont{Kane}},
  \bibinfo{journal}{Nature} \textbf{\bibinfo{volume}{393}},
  \bibinfo{pages}{133} (\bibinfo{year}{1998}).

\bibitem[{\citenamefont{Vrijen et~al.}(2000)\citenamefont{Vrijen, Yablonovitch,
  Wang, Jiang, Balandin, Roychowdhury, Mor, and DiVincenzo}}]{Vrijen}
\bibinfo{author}{\bibfnamefont{R.}~\bibnamefont{Vrijen}},
  \bibinfo{author}{\bibfnamefont{E.}~\bibnamefont{Yablonovitch}},
  \bibinfo{author}{\bibfnamefont{K.}~\bibnamefont{Wang}},
  \bibinfo{author}{\bibfnamefont{H.~W.} \bibnamefont{Jiang}},
  \bibinfo{author}{\bibfnamefont{A.}~\bibnamefont{Balandin}},
  \bibinfo{author}{\bibfnamefont{V.}~\bibnamefont{Roychowdhury}},
  \bibinfo{author}{\bibfnamefont{T.}~\bibnamefont{Mor}}, \bibnamefont{and}
  \bibinfo{author}{\bibfnamefont{D.}~\bibnamefont{DiVincenzo}},
  \bibinfo{journal}{Phys. Rev. A} \textbf{\bibinfo{volume}{62}},
  \bibinfo{pages}{012306} (\bibinfo{year}{2000}).

\bibitem[{\citenamefont{Imamoglu et~al.}(1999)\citenamefont{Imamoglu,
  Awschalom, Burkard, DiVincenzo, Loss, Sherwin, and Small}}]{Imamoglu99}
\bibinfo{author}{\bibfnamefont{A.}~\bibnamefont{Imamoglu}},
  \bibinfo{author}{\bibfnamefont{D.~D.} \bibnamefont{Awschalom}},
  \bibinfo{author}{\bibfnamefont{G.}~\bibnamefont{Burkard}},
  \bibinfo{author}{\bibfnamefont{D.~P.} \bibnamefont{DiVincenzo}},
  \bibinfo{author}{\bibfnamefont{D.}~\bibnamefont{Loss}},
  \bibinfo{author}{\bibfnamefont{M.}~\bibnamefont{Sherwin}}, \bibnamefont{and}
  \bibinfo{author}{\bibfnamefont{A.}~\bibnamefont{Small}},
  \bibinfo{journal}{Phys. Rev. Lett.} \textbf{\bibinfo{volume}{83}},
  \bibinfo{pages}{4204} (\bibinfo{year}{1999}).

\bibitem[{\citenamefont{Kane}(2000)}]{Kane1}
\bibinfo{author}{\bibfnamefont{B.~E.} \bibnamefont{Kane}},
  \bibinfo{journal}{Fortschr. Phys.} \textbf{\bibinfo{volume}{48}},
  \bibinfo{pages}{1023} (\bibinfo{year}{2000}).

\bibitem[{\citenamefont{Larionov et~al.}(2001)\citenamefont{Larionov,
  Fedichkin, and Valiev}}]{larionov01}
\bibinfo{author}{\bibfnamefont{A.~A.} \bibnamefont{Larionov}},
  \bibinfo{author}{\bibfnamefont{L.~E.} \bibnamefont{Fedichkin}},
  \bibnamefont{and} \bibinfo{author}{\bibfnamefont{K.~A.}
  \bibnamefont{Valiev}}, \bibinfo{journal}{Nanotechnology}
  \textbf{\bibinfo{volume}{12}}, \bibinfo{pages}{536} (\bibinfo{year}{2001}).

\bibitem[{\citenamefont{Skinner et~al.}(2003)\citenamefont{Skinner, Davenport,
  and Kane}}]{skinner03}
\bibinfo{author}{\bibfnamefont{A.~J.} \bibnamefont{Skinner}},
  \bibinfo{author}{\bibfnamefont{M.~E.} \bibnamefont{Davenport}},
  \bibnamefont{and} \bibinfo{author}{\bibfnamefont{B.~E.} \bibnamefont{Kane}},
  \bibinfo{journal}{Phys. Rev. Lett.} \textbf{\bibinfo{volume}{90}},
  \bibinfo{pages}{087901} (\bibinfo{year}{2003}).

\bibitem[{\citenamefont{Hu and {Das Sarma}}(2003)}]{overview}
\bibinfo{author}{\bibfnamefont{X.}~\bibnamefont{Hu}} \bibnamefont{and}
  \bibinfo{author}{\bibfnamefont{S.}~\bibnamefont{{Das Sarma}}},
  \bibinfo{journal}{Phys. Stat. Sol. (b)} \textbf{\bibinfo{volume}{238}},
  \bibinfo{pages}{360} (\bibinfo{year}{2003}).

\bibitem[{\citenamefont{Rugar et~al.}(2004)\citenamefont{Rugar, Budakian,
  Mamin, and Chui}}]{Rugar04}
\bibinfo{author}{\bibfnamefont{D.}~\bibnamefont{Rugar}},
  \bibinfo{author}{\bibfnamefont{R.}~\bibnamefont{Budakian}},
  \bibinfo{author}{\bibfnamefont{H.~J.} \bibnamefont{Mamin}}, \bibnamefont{and}
  \bibinfo{author}{\bibfnamefont{B.~W.} \bibnamefont{Chui}},
  \bibinfo{journal}{Nature} \textbf{\bibinfo{volume}{430}},
  \bibinfo{pages}{329} (\bibinfo{year}{2004}).

\bibitem[{\citenamefont{Elzerman et~al.}(2004)\citenamefont{Elzerman, Hanson,
  van Beveren, Witkamp, Vandersypen, and Kouwenhoven}}]{Elzerman}
\bibinfo{author}{\bibfnamefont{J.}~\bibnamefont{Elzerman}},
  \bibinfo{author}{\bibfnamefont{R.}~\bibnamefont{Hanson}},
  \bibinfo{author}{\bibfnamefont{L.~H.~W.} \bibnamefont{van Beveren}},
  \bibinfo{author}{\bibfnamefont{B.}~\bibnamefont{Witkamp}},
  \bibinfo{author}{\bibfnamefont{L.~M.~K.} \bibnamefont{Vandersypen}},
  \bibnamefont{and} \bibinfo{author}{\bibfnamefont{L.~P.}
  \bibnamefont{Kouwenhoven}}, \bibinfo{journal}{Nature}
  \textbf{\bibinfo{volume}{430}}, \bibinfo{pages}{431} (\bibinfo{year}{2004}).

\bibitem[{\citenamefont{Makhlin et~al.}(2000)\citenamefont{Makhlin, Sch{\"o}n,
  and Schnirman}}]{Schon_RMP}
\bibinfo{author}{\bibfnamefont{Y.}~\bibnamefont{Makhlin}},
  \bibinfo{author}{\bibfnamefont{G.}~\bibnamefont{Sch{\"o}n}},
  \bibnamefont{and}
  \bibinfo{author}{\bibfnamefont{A.}~\bibnamefont{Schnirman}},
  \bibinfo{journal}{Rev. Mod. Phys.} \textbf{\bibinfo{volume}{73}},
  \bibinfo{pages}{357} (\bibinfo{year}{2000}).

\bibitem[{\citenamefont{Nakamura et~al.}(1999)\citenamefont{Nakamura, Pashkin,
  and Tsai}}]{Nakamura0}
\bibinfo{author}{\bibfnamefont{Y.}~\bibnamefont{Nakamura}},
  \bibinfo{author}{\bibfnamefont{Y.~A.} \bibnamefont{Pashkin}},
  \bibnamefont{and} \bibinfo{author}{\bibfnamefont{J.~S.} \bibnamefont{Tsai}},
  \bibinfo{journal}{Nature} \textbf{\bibinfo{volume}{398}},
  \bibinfo{pages}{786} (\bibinfo{year}{1999}).

\bibitem[{\citenamefont{Barenco et~al.}(1995)\citenamefont{Barenco, Deutsch,
  Ekert, and Jozsa}}]{Barenco}
\bibinfo{author}{\bibfnamefont{A.}~\bibnamefont{Barenco}},
  \bibinfo{author}{\bibfnamefont{D.}~\bibnamefont{Deutsch}},
  \bibinfo{author}{\bibfnamefont{A.}~\bibnamefont{Ekert}}, \bibnamefont{and}
  \bibinfo{author}{\bibfnamefont{R.}~\bibnamefont{Jozsa}},
  \bibinfo{journal}{Phys. Rev. Lett.} \textbf{\bibinfo{volume}{74}},
  \bibinfo{pages}{4083} (\bibinfo{year}{1995}).

\bibitem[{\citenamefont{Ekert and Jozsa}(1996)}]{Ekert}
\bibinfo{author}{\bibfnamefont{A.~K.} \bibnamefont{Ekert}} \bibnamefont{and}
  \bibinfo{author}{\bibfnamefont{R.}~\bibnamefont{Jozsa}},
  \bibinfo{journal}{Rev. Mod. Phys.} \textbf{\bibinfo{volume}{68}},
  \bibinfo{pages}{733} (\bibinfo{year}{1996}).

\bibitem[{\citenamefont{Tanamoto}(2000)}]{Tanamoto}
\bibinfo{author}{\bibfnamefont{T.}~\bibnamefont{Tanamoto}},
  \bibinfo{journal}{Phys. Rev. A} \textbf{\bibinfo{volume}{61}},
  \bibinfo{pages}{022305} (\bibinfo{year}{2000}).

\bibitem[{\citenamefont{Sherwin et~al.}(1999)\citenamefont{Sherwin, Imamoglu,
  and Montroy}}]{Sherwin}
\bibinfo{author}{\bibfnamefont{M.~S.} \bibnamefont{Sherwin}},
  \bibinfo{author}{\bibfnamefont{A.}~\bibnamefont{Imamoglu}}, \bibnamefont{and}
  \bibinfo{author}{\bibfnamefont{T.}~\bibnamefont{Montroy}},
  \bibinfo{journal}{Phys. Rev. A} \textbf{\bibinfo{volume}{60}},
  \bibinfo{pages}{3508} (\bibinfo{year}{1999}).

\bibitem[{\citenamefont{Hollenberg
  et~al.}(2004{\natexlab{a}})\citenamefont{Hollenberg, Dzurak, Wellard,
  Hamilton, Reilly, Milburn, and Clark}}]{hollenberg1}
\bibinfo{author}{\bibfnamefont{L.~C.~L.} \bibnamefont{Hollenberg}},
  \bibinfo{author}{\bibfnamefont{A.~S.} \bibnamefont{Dzurak}},
  \bibinfo{author}{\bibfnamefont{C.~J.} \bibnamefont{Wellard}},
  \bibinfo{author}{\bibfnamefont{A.~R.} \bibnamefont{Hamilton}},
  \bibinfo{author}{\bibfnamefont{D.~J.} \bibnamefont{Reilly}},
  \bibinfo{author}{\bibfnamefont{G.~J.} \bibnamefont{Milburn}},
  \bibnamefont{and} \bibinfo{author}{\bibfnamefont{R.~G.} \bibnamefont{Clark}},
  \bibinfo{journal}{Phys. Rev. B} \textbf{\bibinfo{volume}{69}},
  \bibinfo{pages}{113301} (\bibinfo{year}{2004}{\natexlab{a}}).

\bibitem[{\citenamefont{Hollenberg
  et~al.}(2004{\natexlab{b}})\citenamefont{Hollenberg, Wellard, Pakes, and
  Fowler}}]{hollenberg2}
\bibinfo{author}{\bibfnamefont{L.~C.~L.} \bibnamefont{Hollenberg}},
  \bibinfo{author}{\bibfnamefont{C.~J.} \bibnamefont{Wellard}},
  \bibinfo{author}{\bibfnamefont{C.~I.} \bibnamefont{Pakes}}, \bibnamefont{and}
  \bibinfo{author}{\bibfnamefont{A.~G.} \bibnamefont{Fowler}},
  \bibinfo{journal}{Phys. Rev. B} \textbf{\bibinfo{volume}{69}},
  \bibinfo{pages}{233301} (\bibinfo{year}{2004}{\natexlab{b}}).

\bibitem[{\citenamefont{Koiller
  et~al.}(2002{\natexlab{a}})\citenamefont{Koiller, Hu, and {Das
  Sarma}}}]{KHD1}
\bibinfo{author}{\bibfnamefont{B.}~\bibnamefont{Koiller}},
  \bibinfo{author}{\bibfnamefont{X.}~\bibnamefont{Hu}}, \bibnamefont{and}
  \bibinfo{author}{\bibfnamefont{S.}~\bibnamefont{{Das Sarma}}},
  \bibinfo{journal}{Phys. Rev. Lett,} \textbf{\bibinfo{volume}{88}},
  \bibinfo{pages}{027903} (\bibinfo{year}{2002}{\natexlab{a}}).

\bibitem[{\citenamefont{Koiller
  et~al.}(2002{\natexlab{b}})\citenamefont{Koiller, Hu, and {Das
  Sarma}}}]{KHD2}
\bibinfo{author}{\bibfnamefont{B.}~\bibnamefont{Koiller}},
  \bibinfo{author}{\bibfnamefont{X.}~\bibnamefont{Hu}}, \bibnamefont{and}
  \bibinfo{author}{\bibfnamefont{S.}~\bibnamefont{{Das Sarma}}},
  \bibinfo{journal}{Phys. Rev. B} \textbf{\bibinfo{volume}{66}},
  \bibinfo{pages}{115201} (\bibinfo{year}{2002}{\natexlab{b}}).

\bibitem[{\citenamefont{Cole et~al.}(2000)\citenamefont{Cole, Williams, King,
  Sherwin, and Stanley}}]{Cole00}
\bibinfo{author}{\bibfnamefont{B.~E.} \bibnamefont{Cole}},
  \bibinfo{author}{\bibfnamefont{J.~B.} \bibnamefont{Williams}},
  \bibinfo{author}{\bibfnamefont{B.~T.} \bibnamefont{King}},
  \bibinfo{author}{\bibfnamefont{M.~S.} \bibnamefont{Sherwin}},
  \bibnamefont{and} \bibinfo{author}{\bibfnamefont{C.~R.}
  \bibnamefont{Stanley}}, \bibinfo{journal}{Nature}
  \textbf{\bibinfo{volume}{410}}, \bibinfo{pages}{60} (\bibinfo{year}{2000}).

\bibitem[{\citenamefont{Hayashi et~al.}(2003)\citenamefont{Hayashi, Fujusawa,
  Cheong, Jeong, and Hirayama}}]{Hayashi}
\bibinfo{author}{\bibfnamefont{T.}~\bibnamefont{Hayashi}},
  \bibinfo{author}{\bibfnamefont{T.}~\bibnamefont{Fujusawa}},
  \bibinfo{author}{\bibfnamefont{H.~D.} \bibnamefont{Cheong}},
  \bibinfo{author}{\bibfnamefont{Y.~H.} \bibnamefont{Jeong}}, \bibnamefont{and}
  \bibinfo{author}{\bibfnamefont{Y.}~\bibnamefont{Hirayama}},
  \bibinfo{journal}{Phys. Rev. Lett.} \textbf{\bibinfo{volume}{91}},
  \bibinfo{pages}{226804} (\bibinfo{year}{2003}).

\bibitem[{\citenamefont{Slater}(1963)}]{slater}
\bibinfo{author}{\bibfnamefont{J.~C.} \bibnamefont{Slater}},
  \emph{\bibinfo{title}{Quantum Theory of Molecules and Solids}},
  vol.~\bibinfo{volume}{1} (\bibinfo{publisher}{McGraw-Hill, New York},
  \bibinfo{year}{1963}).

\bibitem[{\citenamefont{Kohn}(1957)}]{Kohn}
\bibinfo{author}{\bibfnamefont{W.}~\bibnamefont{Kohn}},
  \emph{\bibinfo{title}{Solid State Physics Series}}, vol.~\bibinfo{volume}{5}
  (\bibinfo{publisher}{Academic Press}, \bibinfo{year}{1957}),
  \bibinfo{note}{edited by F. Seitz and D. Turnbull, p.257, and references
  therein}.

\bibitem[{foo()}]{footvalley}
\bibinfo{note}{Here we consider unstrained Si, thus the nondegenerate $A_1$
  ground state corresponds to all $\alpha_\mu = 1/\sqrt{6}$. The valley
  populations change\cite{KHD2} when strain is applied to the sample}.

\bibitem[{\citenamefont{Koiller et~al.}(2004)\citenamefont{Koiller, Capaz, Hu,
  and {Das Sarma}}}]{KCHD}
\bibinfo{author}{\bibfnamefont{B.}~\bibnamefont{Koiller}},
  \bibinfo{author}{\bibfnamefont{R.~B.} \bibnamefont{Capaz}},
  \bibinfo{author}{\bibfnamefont{X.}~\bibnamefont{Hu}}, \bibnamefont{and}
  \bibinfo{author}{\bibfnamefont{S.}~\bibnamefont{{Das Sarma}}},
  \bibinfo{journal}{Phys. Rev. B} \textbf{\bibinfo{volume}{70}},
  \bibinfo{pages}{115207} (\bibinfo{year}{2004}).

\bibitem[{\citenamefont{Schofield et~al.}(2003)\citenamefont{Schofield, Curson,
  Simmons, Rue\ss, Hallam, Oberbeck, and Clark}}]{encapsulation}
\bibinfo{author}{\bibfnamefont{S.~R.} \bibnamefont{Schofield}},
  \bibinfo{author}{\bibfnamefont{N.~J.} \bibnamefont{Curson}},
  \bibinfo{author}{\bibfnamefont{M.~Y.} \bibnamefont{Simmons}},
  \bibinfo{author}{\bibfnamefont{F.~J.} \bibnamefont{Rue\ss}},
  \bibinfo{author}{\bibfnamefont{T.}~\bibnamefont{Hallam}},
  \bibinfo{author}{\bibfnamefont{L.}~\bibnamefont{Oberbeck}}, \bibnamefont{and}
  \bibinfo{author}{\bibfnamefont{R.~G.} \bibnamefont{Clark}},
  \bibinfo{journal}{Phys. Rev. Lett.} \textbf{\bibinfo{volume}{91}},
  \bibinfo{pages}{136104} (\bibinfo{year}{2003}).

\bibitem[{\citenamefont{Nielsen and Chuang}(2000)}]{NielsenChuang}
\bibinfo{author}{\bibfnamefont{M.~A.} \bibnamefont{Nielsen}} \bibnamefont{and}
  \bibinfo{author}{\bibfnamefont{I.~L.} \bibnamefont{Chuang}},
  \emph{\bibinfo{title}{Quantum Computation and Quantum Information}}
  (\bibinfo{publisher}{Cambridge, Cambridge}, \bibinfo{year}{2000}).

\bibitem[{\citenamefont{Barrett and Milburn}(2003)}]{Barrett}
\bibinfo{author}{\bibfnamefont{S.~D.} \bibnamefont{Barrett}} \bibnamefont{and}
  \bibinfo{author}{\bibfnamefont{G.~J.} \bibnamefont{Milburn}},
  \bibinfo{journal}{Phys. Rev. B} \textbf{\bibinfo{volume}{68}},
  \bibinfo{pages}{155307} (\bibinfo{year}{2003}).

\bibitem[{\citenamefont{Mahan}(1981)}]{Mahan}
\bibinfo{author}{\bibfnamefont{G.~D.} \bibnamefont{Mahan}},
  \emph{\bibinfo{title}{Many-Particle Physics}} (\bibinfo{publisher}{Plenum,
  New York}, \bibinfo{year}{1981}).

\bibitem[{\citenamefont{Bastard}(1988)}]{Bastard}
\bibinfo{author}{\bibfnamefont{G.}~\bibnamefont{Bastard}},
  \emph{\bibinfo{title}{Wave mechanics applied to semiconductor
  heterostructures}} (\bibinfo{publisher}{Halsted, New York},
  \bibinfo{year}{1988}).

\bibitem[{\citenamefont{Fedichkin and Fedorov}(2004)}]{Fedichkin}
\bibinfo{author}{\bibfnamefont{L.}~\bibnamefont{Fedichkin}} \bibnamefont{and}
  \bibinfo{author}{\bibfnamefont{A.}~\bibnamefont{Fedorov}},
  \bibinfo{journal}{Phys. Rev. A} \textbf{\bibinfo{volume}{69}},
  \bibinfo{pages}{032311} (\bibinfo{year}{2004}).

\bibitem[{not()}]{notedephasing}
\bibinfo{note}{Calculation of pure dephasing rate is not constrained by energy
  conservation because pure dephasing processes do not involve energy
  transfer.\cite{Duan} Therefore higher-energy phonons should also contribute
  to the total dephasing rate (although electron-phonon coupling matrix
  elements becomes smaller as $|{\bf q}|$ increases), so that validity of the
  envelope function simplifications employed in the present study becomes
  questionable and a direct numerical calculation may be necessary}.

\bibitem[{\citenamefont{Hu and {Das Sarma}}(2000)}]{HD}
\bibinfo{author}{\bibfnamefont{X.}~\bibnamefont{Hu}} \bibnamefont{and}
  \bibinfo{author}{\bibfnamefont{S.}~\bibnamefont{{Das Sarma}}},
  \bibinfo{journal}{Phys. Rev. A} \textbf{\bibinfo{volume}{61}},
  \bibinfo{pages}{062301} (\bibinfo{year}{2000}).

\bibitem[{\citenamefont{Nakamura et~al.}(2002)\citenamefont{Nakamura, Pashkin,
  Yamamoto, and Tsai}}]{Nakamura}
\bibinfo{author}{\bibfnamefont{Y.}~\bibnamefont{Nakamura}},
  \bibinfo{author}{\bibfnamefont{Y.~A.} \bibnamefont{Pashkin}},
  \bibinfo{author}{\bibfnamefont{T.}~\bibnamefont{Yamamoto}}, \bibnamefont{and}
  \bibinfo{author}{\bibfnamefont{J.~S.} \bibnamefont{Tsai}},
  \bibinfo{journal}{Phys. Rev. Lett.} \textbf{\bibinfo{volume}{88}},
  \bibinfo{pages}{047901} (\bibinfo{year}{2002}).

\bibitem[{\citenamefont{Itakura and Tokura}(2003)}]{Itakura}
\bibinfo{author}{\bibfnamefont{T.}~\bibnamefont{Itakura}} \bibnamefont{and}
  \bibinfo{author}{\bibfnamefont{Y.}~\bibnamefont{Tokura}},
  \bibinfo{journal}{Phys. Rev. B} \textbf{\bibinfo{volume}{67}},
  \bibinfo{pages}{195320} (\bibinfo{year}{2003}).

\bibitem[{\citenamefont{Duan and Guo}(1998)}]{Duan}
\bibinfo{author}{\bibfnamefont{L.~M.} \bibnamefont{Duan}} \bibnamefont{and}
  \bibinfo{author}{\bibfnamefont{G.~C.} \bibnamefont{Guo}},
  \bibinfo{journal}{Phys. Rev. A} \textbf{\bibinfo{volume}{57}},
  \bibinfo{pages}{737} (\bibinfo{year}{1998}).

\end{thebibliography}

\end{document}